\documentclass{emulateapj}

\newcommand\simless{{\thinspace \rlap{\raise 0.5ex\hbox{$\scriptstyle
 {<}$}} {\lower 0.3ex\hbox{$\scriptstyle {\sim}$}} \thinspace }}  
\newcommand\simgreat{{\thinspace \rlap{\raise
 0.5ex\hbox{$\scriptstyle {>}$}} {\lower 0.3ex\hbox{$\scriptstyle
 {\sim}$}} \thinspace }} 
\newcommand\msun{\, \rm M_\odot}

\newcommand\ergs{{\, \rm erg\,s^{-1}}} 
\newcommand{\be}{\begin{equation}}
\newcommand{\ba}{\begin{eqnarray}}
\newcommand{\ee}{\end{equation}}
\newcommand{\ea}{\end{eqnarray}}

\slugcomment{Submitted to ApJ}

\begin{document}

\title{The trail of discrete X-ray sources \break in the early-type galaxy
  NGC~4261: \break anisotropy in the globular cluster distribution?}

\author{Lea Giordano\altaffilmark{1}, Luca Cortese\altaffilmark{1,4}, Ginevra
  Trinchieri\altaffilmark{2}, Anna Wolter\altaffilmark{2}, Monica
  Colpi\altaffilmark{1}, Giuseppe Gavazzi\altaffilmark{1}, Lucio
  Mayer\altaffilmark{3}}

\altaffiltext{1}{Universit\`{a} degli Studi di Milano Bicocca, Piazza della
  Scienza 3, 20126 Milano, Italy} \altaffiltext{2}{Osservatorio Astronomico di
  Brera, Via Brera 28, 20121 Milano, Italy} \altaffiltext{3}{Institute of
  Theoretical Physics, University of Z\"{u}rich, Winterurestrasse 190, 8057
  Z\"{u}rich, Switzerland} \altaffiltext{4}{Laboratoire d'Astrophysique de
  Marseille, BP8, Traverse du Siphon, 13376 Marseille Cedex 12, France}

\begin{abstract}
  {\it Chandra} images of the elliptical galaxy NGC~4261 have revealed
  an anisotropy in the spatial distribution of the off-nuclear X-ray
  sources, interpreted by Zezas et al. as evidence of an association
  with a young stellar population.  Our independent analysis of
  archive X-ray ({\it Chandra}) and optical (INT and {\it HST})
  observations confirms the anisotropy of the X-ray sources but
  conducts to a different interpretation for their origin.  We find
  that nearly half of the X-ray sources are associated to a globular
  cluster (optical counterpart) suggesting that they are accreting
  low-mass X-ray binaries.  Where color index information is
  available, the X-ray sources are found to reside in red (metal-rich)
  systems.  The luminosity function of the X-ray sources is also
  consistent with the one drawn from a population of Low Mass X-ray
  Binaries.  We further investigate the properties of the sample of
  point-like sources obtained from archival optical images that we
  suggest are good globular cluster candidates and for which we find
  that the projected spatial distribution is non-homogeneous. In
  addition, we show that the distributions of the optical and X-ray
  populations are very similar, which leads us to conclude that the
  spatial anisotropy of the X-ray sources in NGC~4261 is mostly a
  reflection of the anisotropy of the globular cluster population.
\end{abstract}


\section{Introduction}
\setcounter{footnote}{0}

NGC~4261 is an early-type galaxy in the Virgo West Cloud, at the
distance of 32 Mpc (\citealt{Gavazzi99}). It belongs to a group of
galaxies (\citealt{Garcia93, Nolthenius93}) that does not show
prominent sign of gravitational interactions.  The H-band luminosity
of NGC~4261 is $7.9\times 10^{10} \rm~L_{\odot}$ corresponding to a
dynamical mass of $9.5\times 10^{10}\msun$ (using a mass to light
ratio of about 1.2, from \citealt{Zibetti02}), ranking this galaxy
among the most massive galaxies in Virgo. The optical spectrum is
consistent with a dominant population of old stars ($\sim 11$ Gyr,
\citealt{Gavazzi02}) and the light distribution is smooth with no
evidence of sub-structure (\citealt{Schweizer92};
\citealt{Colbert01}).  NGC~4261 is a FRI radio galaxy (3C 270) showing
two radio jets emanating from a nucleus hosting a supermassive black
hole with a mass of about 5 $\times$ 10$^8$ M$_{\odot}$
(\citealt{Ferrarese96}) surrounded by a dusty torus and a nuclear disk
(\citealt{Jaffe96}; \citealt{Ferrarese96}). The X-ray nuclear power
from the underlying AGN has been well studied by {\it XMM} and {\it
Chandra}: the X-ray luminosity is L$_X \sim 5 \times 10^{41}$ erg
s$^{-1}$, and the X-ray spectrum is a power law absorbed by a column
density N$_{\rm {H}}\sim 5\times 10^{20}$ cm$^{-2}$. The source
appears to be embedded in diffuse hot gas ($k_BT \sim 0.6$ keV) and
shows low amplitude, rapid variability (\citealt{Chiaberge03};
\citealt{Sambruna03}; \citealt{Gliozzi03}).

Thanks to the high spatial resolution of {\it Chandra},
\citealt{Zezas03} (ZHFM03 hereafter) have discovered the presence of
about sixty bright off-nuclear X-ray sources in NGC~4261, which stand
out for their large-scale anisotropic pattern.  The anisotropy extends
over an angle of $\sim 4'$, corresponding to a linear dimension of
$\sim 40$ kpc ($\sim 5$ effective core radii large, since $r_e=4.6$
kpc for NGC~4261; \citealt{Gavazzi03}). ZHFM03 interpret this
anisotropy as evidence that the X-ray sources are associated with
young stellar population, possibly formed in a recent episode of star
formation triggered by some dynamical agent along tidal tails.  In
spite of any optical evidence of either a young population (the
population in the nuclear regions is as old as 15 Gyrs; see
\citealt{Gavazzi03} and \citealt{Goudfrooij94}) or of recent merging
events, ZHFM03 suggest a high mass binary origin for the majority of
the sources, based on the significantly higher efficiency of the
formation of High Mass X-ray Binaries relative to the Low Mass ones.
This interpretation makes NGC~4261 a rather unique example, in clear
contrast with {\it Chandra} observations of other early-type galaxies,
where the X-ray sources are generally distributed nearly
isotropically, trace the dominant (old) stellar population, and are
believed to belong to the Low Mass X-ray Binary population (LMXB,
e.g. \citealt{Verbunt84}; \citealt{Sarazin03}; \citealt{Kim04}).  In
addition, a significant fraction ($\sim$ 20\% -70\% ) of these LMXBs
is associated to globular clusters (GCs; \citealt{Sarazin03};
\citealt{Angelini01}; \citealt{Kundu02}; \citealt{Maccarone03};
\citealt{Minniti04}), with a preference to GCs with redder color
index, i.e., clusters that are more metal rich (\citealt{Cote98};
\citealt{Sarazin03}; \citealt{West04}).  In this perspective, the case
of NGC~4261 stands out as a remarkable exception worth of further
study.

We have therefore chosen to study its case again, in order to deepen
our understanding of the the nature of the X-ray sources in external
galaxies.  For this reason we have used archive observations of {\it
Chandra}, of the Isaac Newton Telescope\footnote{Roque de Los
Muchachos Observatory, La Palma, Spain
(\texttt{http://www.ing.iac.es/})} (INT), and of the Hubble Space
Telescope ({\it HST}) (see Table 1 for a summary of the dataset used)
to search for optical counterparts of the X-ray sources, and to study
their relation with the GC population of NGC~4261.

The outline of the paper is as follows.  In $\S 2$ we describe the
data reduction, in the X-ray and optical bands.  In $\S 3$ we proceed
on studying the properties of the optical point-like sources, and we
cross-correlate the optical data with the X-ray data. We then derive
the luminosity function of the X-ray source population. In $\S 4$ we
investigate on the azimuthal anisotropy of the X-ray sources, and on
that of the underlying GC candidate population.  In the context of our
new findings we discuss in $\S 5$ the nature of the sources and their
potential in tracing the history of assembly of NGC~4261 in the Virgo
cluster.

\section{Data reduction}

\subsection{X-ray data}

\begin{figure}
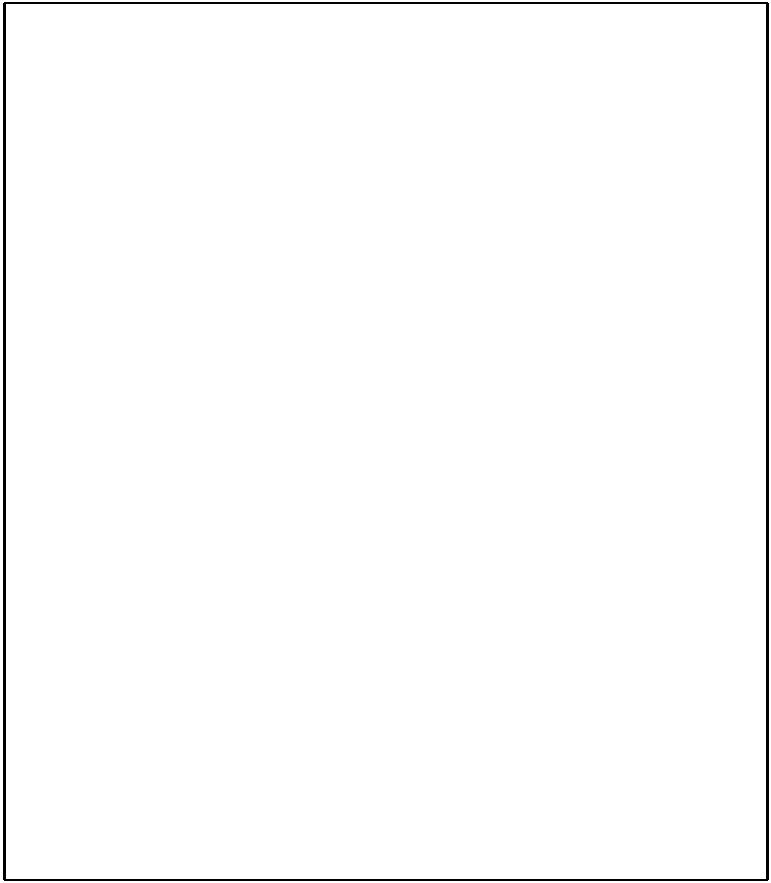

\centering 
\begin{tabular}{|p{10.0cm}|}
\hline
\\ \\ \\ \\ \\ \\ \\ \\ \\ \\ \\ \\ \\ \\ \\ \\ \\
\\ \\ \\ \\ \\ \\ \\ \\ \\ \\ \\ \\ \\ \\ \\ \\ \\
\hline
\end{tabular}
\caption{R band image of NGC~4261 from the INT telescope.  The active
  field of the {\it Chandra} ACIS-S3 CCD and {\it HST} WFPC2 fields
  are superposed in black and white, respectively.  }
\label{fig1}
\end{figure}

\begin{figure}
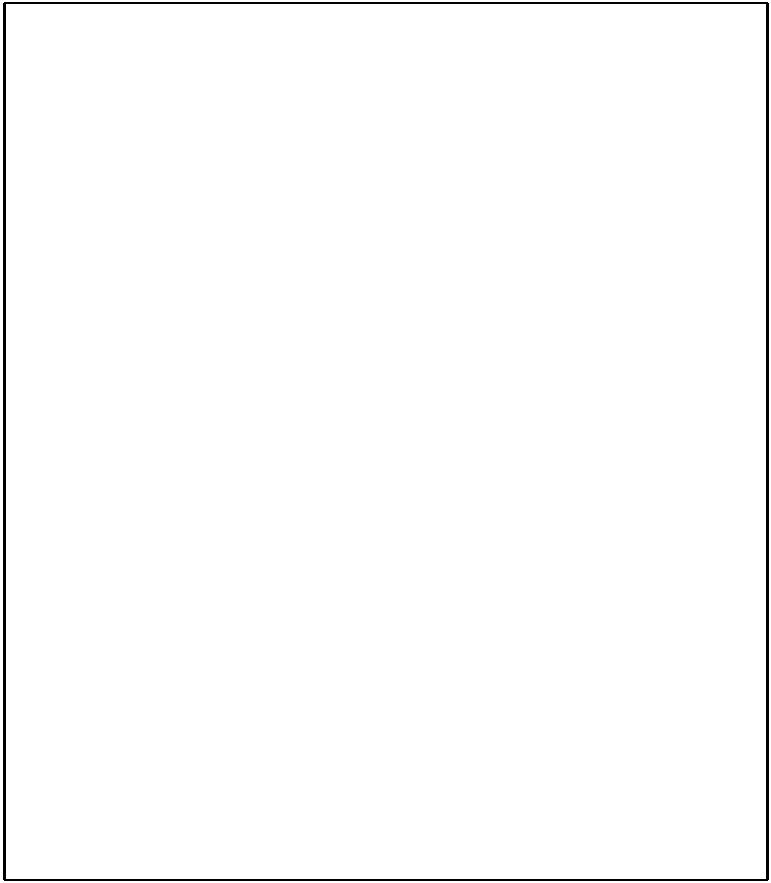

\centering
\begin{tabular}{|p{10.0cm}|}
\hline
\\ \\ \\ \\ \\ \\ \\ \\ \\ \\ \\ \\ \\ \\ \\ \\ \\
\\ \\ \\ \\ \\ \\ \\ \\ \\ \\ \\ \\ \\ \\ \\ \\ \\
\hline
\end{tabular}
\caption{{\it Chandra} S3 image of NGC~4261. Open squares correspond
  to X-ray sources with optical counterpart while open circles to
  sources without optical counterpart.  Excluded from the analysis are
  the X-ray sources (diamonds) that fall in the region where the model
  subtraction is too noisy, and source no. 54 located outside the
  galaxy radius. The black cross corresponds to the AGN.  ``Plus''
  symbols correspond to background/foreground galaxies.}
\label{fig2}
\end{figure}

NGC~4261 was observed by {\it Chandra} \citep{Weisskopf00} with the
ACIS-S3 CCD on 2000 May 6 for a total exposure of 37,210 s (see Table
\ref{tab:obs} for details). The data were collected in 1/2 sub-array
mode to avoid the pile-up of the central AGN source. The active field
of view is shown in Figure \ref{fig1} over-plotted on an optical image
of the galaxy.  The data reduction was performed with the package
\texttt{CIAO}\footnote{{\it Chandra} Interactive Analysis of
Observations (CIAO), \texttt{http://cxc.harvard.edu/ciao/}} v. 3.0.2
and \texttt{CALDB} v. 2.2. Data were cleaned for high background
periods due to particle induced flares and the resulting exposure time
is of $28.3$ ks.  Individual sources were detected using a
wavelet-detection algorithm (\emph{wavdetect} in \texttt{CIAO}), with
the same parameters used by ZHFM03, in three energy bands: full band
($0.3 -7.0$ keV), soft band ($0.3 - 2.0$ keV) and hard band ($2.0 -
7.0$ keV).  The catalog that we obtained includes 54 sources detected
above the probability threshold of 10$^{-6}$ that implies $\le$ 1
false detection due to statistical fluctuations.  The positions of the
X-ray sources are shown in Figure \ref{fig2}.  In Table
\ref{tab:sources} we give the source number (column 1), the right
ascension and declination J2000 (column 2 and 3).  Count rates (column
4) are converted into unabsorbed luminosities ($0.2 - 10$ keV)
assuming a power-law model with photon index 1.7 and a Galactic line
of sight column density of N$_{\rm H}$= 5.8 $\times$ 10$^{20}$
cm$^{-2}$ (\citealt{Stark92}).  ZHFM03 do not publish a catalog of
their X-ray sources, so a comparison of the two source lists is not
possible.  However, a visual inspection of their Figure 1 indicate
consistency.  The {\it Chandra} data relative to the central region of
this galaxy has also been analyzed by \cite{Chiaberge03} and
\cite{Gliozzi03}, who have mostly focused on the AGN emission in the
X-ray, radio and optical bands.

In particular \cite{Gliozzi03} detect 5 off-nuclear sources: A, B, C,
E coincide with our sources 38, 33, 31, and 14 respectively, and D,
which we do not detect (our source 19 is the closest, at $\sim 10''$).
Although these authors comment of the fact that they do not find
optical counterparts for any of their sources, they do not comment on
the possible connection of these sources with the jet, in spite of the
evidence that two of them, (A and E) are aligned and at opposite ends
of it.

\subsection{Optical data}

We have used archive observations of INT and {\it HST} WFPC2 for a
systematic search of the optical counterparts of the X-ray sources.

We extracted a deep $r'$ band image from the INT-Wide Field Camera
(WFC) archive, consisting of three exposures of 500 seconds each, for
a total integration time of 1500 seconds.  The image covers the whole
galaxy (the field of view of the INT WFC is of about $0.5$ deg$^2$)
and has a mean seeing of $\sim 1''$ (\citealt{Khosroshahi04}).  The
reduction of the images was carried out using standard tasks in the
\texttt{IRAF} package\footnote {IRAF is the Image Analysis and
Reduction Facility made available to the astronomical community by the
National Optical Astronomy Observatories, which are operated by AURA,
Inc., under contract with the U.S. National Science Foundation. STSDAS
is distributed by the Space Telescope Science Institute, which is
operated by the Association of Universities for Research in Astronomy
(AURA), Inc., under NASA contract NAS 5--26555.}. Bias subtraction and
flat-field normalization were applied to each exposure using the
median of several bias frames and flat-field exposures obtained during
the observations. The three exposures were combined using a median
filter, that automatically removes cosmic rays. Using reference
catalogues of astrometric stars, we obtained a WCS solution for the
whole field having a mean RMS of $\sim$1 arcsec.

{\it HST} WFPC2 observations cover the central portion of the galaxy,
as shown by the white boxes in Figure \ref{fig1}.  The south-east (SE)
field was observed using three different filters: F675W, F547M, F791W
while only the F702W filter is available for the north-west (NW)
field. The exposure times of the four images are given in Table
\ref{tab:obs}.

For our analysis we started from pipelined processed data. For both
{\it HST} WFPC2 and INT, we first fitted ellipses to the isophote of
the galaxy using the task \texttt{ELLIPSE} from the IRAF package, and
subtracted the model from the data. The resulting image was then used
to detect sources with SExtractor (\citealt{SExtractor}).  The last
isophote has a major and minor axis of $4.5$ and $3.2$ arcmin
respectively.  We improve the relative HST astrometry from $\sim 0.7$
to $\sim 0.2$ arcsec comparing the overlap region between the NW and
SE field. On the contrary we did not try to make any improvement in
the relative astrometry between Chandra and INT/HST images.  We note
that, for both HST and INT images, the model subtraction leaves a
quadrupole structure in the central region.  Although this region is
smaller in the HST data, we exclude a wider area of $\sim$0.5 square
arcmin, due to the presence of residuals in the INT data that could
induce spurious results in our comparisons.  This structure was
already noticed by \cite{Colbert01} for this object, and was
attributed to the boxyness of the galaxy.

Using SExtractor we extracted point-like objects from the model
subtracted images of both INT and {\it HST} datasets.  We set a
SExtractor detection threshold of 2.5 $\sigma$ over the local
background. We select as point-like sources objects with the following
properties: (i) a magnitude error smaller than 0.1, (ii) ellipticity
smaller than 0.5, (iii) FWHM between 1 and 4 pixels (0.1-0.4 arcsec)
for {\it HST}, as suggested by \cite{Gebhardt99}, and FWHM between 3
and 10 pixels (1-3 arcsec) for INT.  Note that for the {\it HST}-SE
field we based our selection upon the F675W filter owing to its longer
exposure.  As a final step, we have checked the SExtractor results by
visual inspection, and we have removed objects on or close to the CCDs
edges for the HST images. 

\section{Results}

\subsection{Selection of candidate globular clusters}

\begin{figure}
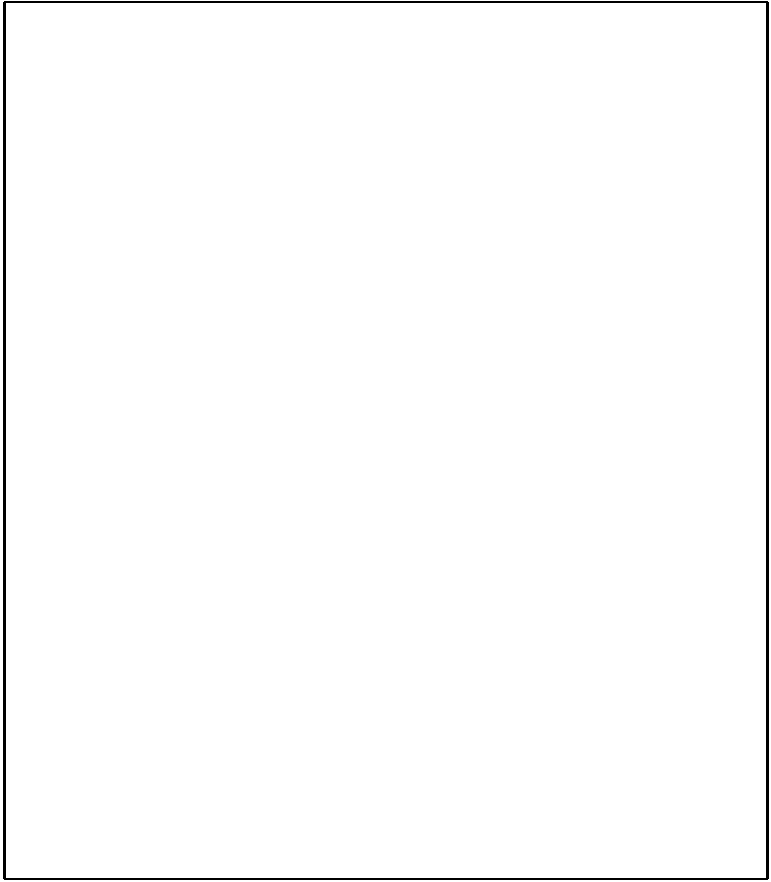

\centering 
\begin{tabular}{|p{10.0cm}|}
\hline
\\ \\ \\ \\ \\ \\ \\ \\ \\ \\ \\ \\ \\ \\ \\ \\ \\
\\ \\ \\ \\ \\ \\ \\ \\ \\ \\ \\ \\ \\ \\ \\ \\ \\
\hline
\end{tabular}
\caption{Histogram showing the difference in the estimate of the V-I
  color between unsubtracted and model subtracted HST images. The
  distribution of the GCs hosting an X-ray source is indicated in
  black.}
\label{fig3}
\end{figure}

\begin{figure}
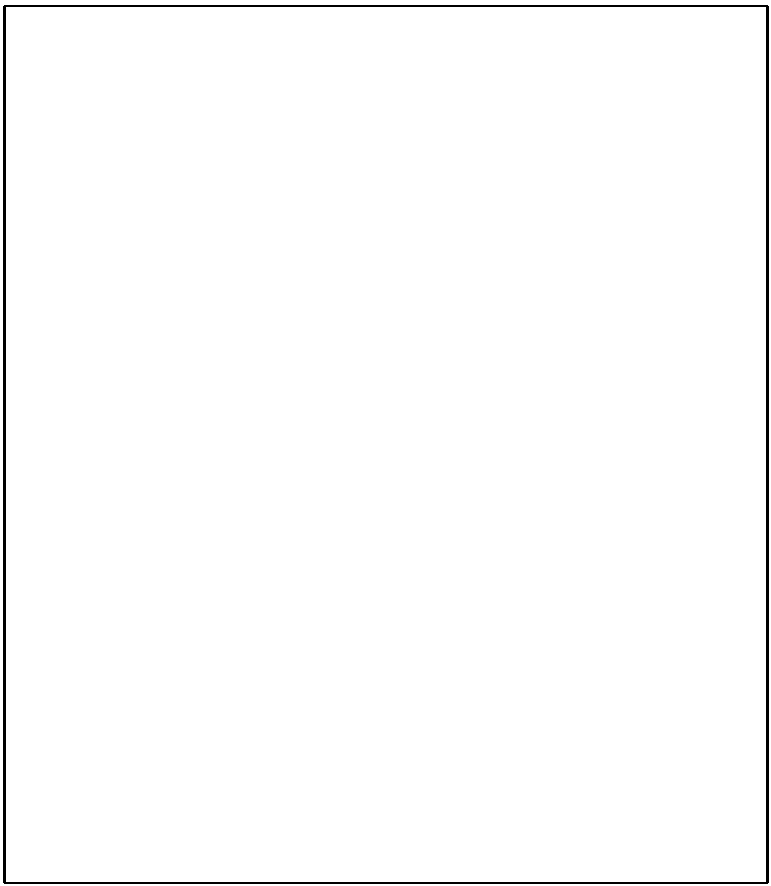

\centering 
\begin{tabular}{|p{10.0cm}|}
\hline
\\ \\ \\ \\ \\ \\ \\ \\ \\ \\ \\ \\ \\ \\ \\ \\ \\
\\ \\ \\ \\ \\ \\ \\ \\ \\ \\ \\ \\ \\ \\ \\ \\ \\
\hline
\end{tabular}
\caption{Left panel: Color index histogram of GCs in {\it HST} SE
  field.  Right panel: Magnitude histogram for the GCs in {\it HST} SE
  field. In both panels the distributions of the GCs hosting an X-ray
  source are superposed in grey. Note that the X-ray sources inhabit
  preferentially metal rich (red), brighter GCs.  }
\label{fig4}
\end{figure}

For the SE field of the {\it HST} observation, we have been able to
estimate the color of the sources presented in $\S 2.2$. The
instrumental magnitudes, provided by SExtractor, were converted in the
Johnson-Cousin V, R and I systems using the synthetic transformations
given in \citealt{Holtzman95}. In order to estimate the photometric
error due to model subtraction we measured the photometry on the
unsubtracted image using local background and apertures from the
subtracted image. The difference in the V-I color of each globular
cluster obtained from the two different estimates is shown in Figure
\ref{fig3}. The distribution appears fairly gaussian with a mean
difference of $-0.003\pm0.07$ mag (see Figure \ref{fig3}).  The total
photometric uncertainty is thus given by the combination of the error
on the measure of GCs magnitudes (given by SExtractor), the error on
the zero-point (taken from \citealp{Holtzman95}) and the uncertainty
due to the model subtraction (0.07 mag, Fig.\ref{fig3}).  This results
in a mean error on the V-I color of $\sim$ 0.11 mag.  We used F547M
and F791W filters to determine the color index (V-I) of the extracted
objects.  We select 325 objects detected in each of the three filters
with V-I color between 0 and 2 (\citealt{Gebhardt99}), that we
consider GC.  Their color index and magnitude distributions shown in
Figure \ref{fig4} are consistent with those of the globular cluster
populations observed in other massive early-type galaxies
(\citealt{Ashman98}; \citealt{Cote98}; \citealt{Kundu01b};
\citealt{Larsen01b}; \citealt{West04}), reinforcing our
definition. The color index distribution spreads from V-I $\sim 0.5$
to $\sim 2$, indicating that both populations of GC (red and blue) are
present.  Thus the unimodal distribution observed in the V-I color
histogram is probably due to the blending of different GCs
populations.

Since the color information is restricted to the {\it HST}-SE field,
we cannot classify as GCs the sources extracted from the other optical
images at the same confidence.  However, we have first considered
sources in the region common to all three datasets ($\sim 1$
arcmin$^2$). The cross-correlation between the three lists indicates
that $\sim~97$\% (37/38) and $\sim 83$\% (49/59) of the sources in the
{\it HST}-NW and INT field respectively coincide with a GC in the {\it
HST}-SE field.  Given this high fraction, we are confident that even
the point-like sources from the INT dataset are good ``GC
candidates'', in spite of the lower resolution of the instrument.  In
what follows, we will use the ``GC candidate'' list (from the INT
observation), that provides an homogeneous set of optical sources over
the whole galaxy.

\begin{figure}
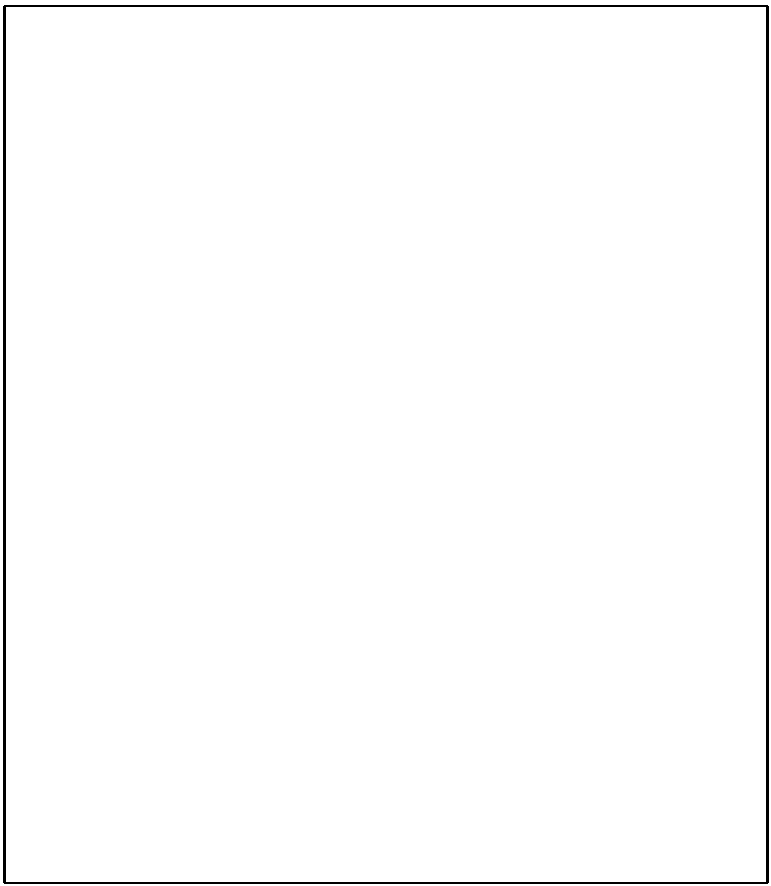

\centering 
\begin{tabular}{|p{10.0cm}|}
\hline
\\ \\ \\ \\ \\ \\ \\ \\ \\ \\ \\ \\ \\ \\ \\ \\ \\
\\ \\ \\ \\ \\ \\ \\ \\ \\ \\ \\ \\ \\ \\ \\ \\ \\
\hline
\end{tabular}
\caption{Adaptive kernel density analysis: Iso-density contours (thin
  line) of the analysis carried on the sample of GC candidates from
  INT are superposed to the iso-density contours (thick line) of the
  GC sample for {\it HST} fields. The first contour represents the 2
  $\sigma$ level.  Black and white regions are excluded from the
  extraction of the optical sources. The two main overdensities
  coincide spatially in both data sets.}
\label{fig5}
\end{figure}

As a final check, we have also compared the sky projected distribution
of the INT and {\it HST} GC candidates. Since {\it HST}-SE and {\it
HST}-NW fields have different exposure times, we restrict the {\it
HST} sample to a magnitude $R$ = $24.5$ (i.e. the magnitude limit of
the shallowest field).  The comparison is done through an adaptive
kernel density analysis. This technique is a two-step procedure which
first applies a pilot smoothing to estimate the local density of GCs,
starting from the list of their positions only, and then uses a
smoothing window variable in size that decreases with increasing local
density.  In this way the statistical noise in low-density regions can
be reduced without smearing out the high density peaks
\citep{Silverman86}.  As shown in Figure \ref{fig5}, the iso-density
contours from the adaptive kernel analysis are in very good agreement,
confirming the reliability of the INT GCs dataset.  What is apparent
in the figure is the peculiar distribution of the GC population, that
shows two main concentrations NE and SW of the nuclear region.  The
presence of the same structures in both datasets ensures that it is
not a feature derived from the poor subtraction in the central region
(the black area in Figure \ref{fig5}) since the equivalent area in the
HST data is negligible.

\subsection{Cross-correlation of X-Ray and Optical data}

\begin{figure}[!ht]
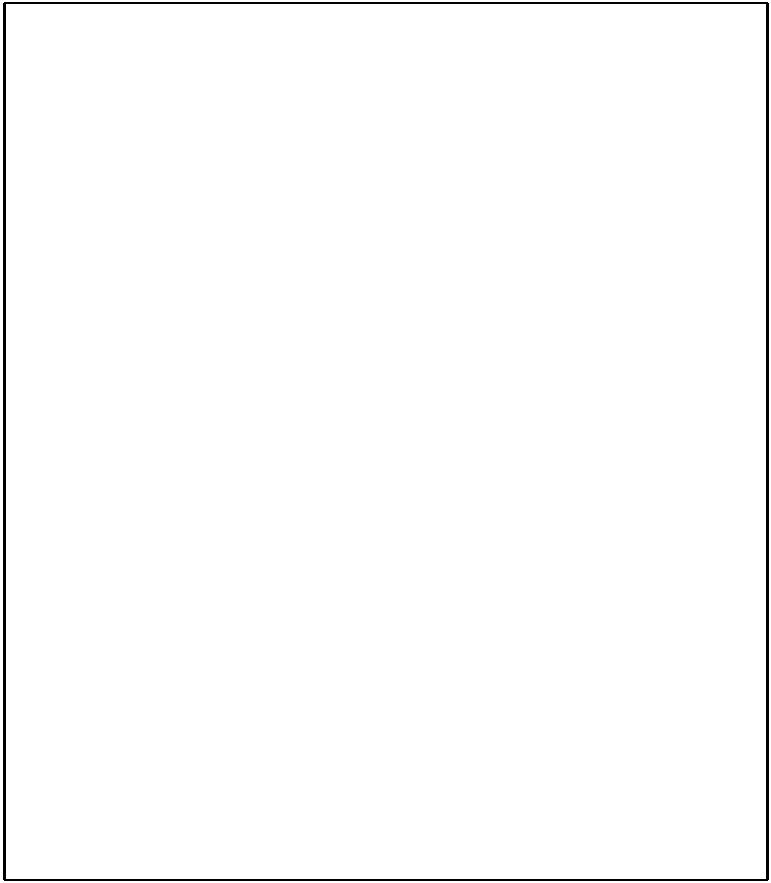

\centering
\begin{tabular}{|p{10.0cm}|}
\hline
\\ \\ \\ \\ \\ \\ \\ \\ \\ \\ \\ \\ \\ \\ \\ \\ \\
\\ \\ \\ \\ \\ \\ \\ \\ \\ \\ \\ \\ \\ \\ \\ \\ \\
\hline
\end{tabular}
\caption{Selected X-ray sources listed in Table \ref{tab:sources} and
  indicated with a star in column 1. The underlying images are
  from the {\it HST} SE (left), {\it HST} NW (middle) and INT (right)
  fields, respectively.  Black circles (of radius equal to 0.7 arcsec
  for {\it HST} and 1 arcsec for INT) indicate the combined
  astrometric uncertainties in the position of each {\it Chandra}
  source.}
\label{fig6}
\end{figure}

In order to investigate the nature of the X-ray sources we have looked
for optical counterparts.  The central source (n.  29) is identified
with the AGN (\citealt{Chiaberge03}; \citealt{Sambruna03};
\citealt{Gliozzi03}), and sources no. 27, 43 and 53 with
background/foreground galaxies.  The positions of the remaining X-ray
sources have been cross-correlated with the GC sample, from the HST
fields, with a matching radius of $0.5$ arcsec, and from the INT
dataset, with a matching radius of $1$ arcsec. These choices take into
account only the intrinsic astrometric uncertainties of the three
datasets (see sections 2.1 and 2.2) and not the relative uncertainties
eventually present between the optical and X-ray observations.  Source
54 falls outside the model subtracted area, while sources no. 19, 31
and 33 fall in the central region excluded from our previous analysis.
We found an optical counterpart for 23 of the 46 discrete X-ray
sources that we assume all belonging to NGC~4261.  The identification
is indicated in Table \ref{tab:sources}.  To calculate the chance
coincidence probability we performed 1000 Monte-Carlo simulations by
shifting the positions of the INT and HST optical samples by a random
offset and performing again the cross-correlation. We adopted a shift
larger than the search radius but small enough so the sources do not
fall outside the distribution of the X-ray sources ($\sim$1.5
arcmin). The probability of finding more than one spurious association
is $\sim$2.5\% and $\sim$0.1\% in the INT and HST sample,
respectively.  In Figure \ref{fig6} we give a few examples that show
the positions of the X-ray sources over-plotted onto the {\it HST} SE,
{\it HST} NW and INT fields.  The figure also demonstrates the good
correspondence between the {\it HST} and INT counterparts, indicating
again that the identifications obtained in the INT field not covered
by {\it HST} are not severely hampered by the lower resolution of the
INT image.  Note that, where the color information is available, we
found that all the X-ray sources inhabit a red (color index V-I $> 1$)
GC, see Figure \ref{fig4}.  The optical identification with GC
candidates for $\sim 50$\% of the X-ray sources suggests a LMXB origin
for them (\citealt{Verbunt04}).

\subsection{Luminosity Function of X-ray sources}

\begin{figure}
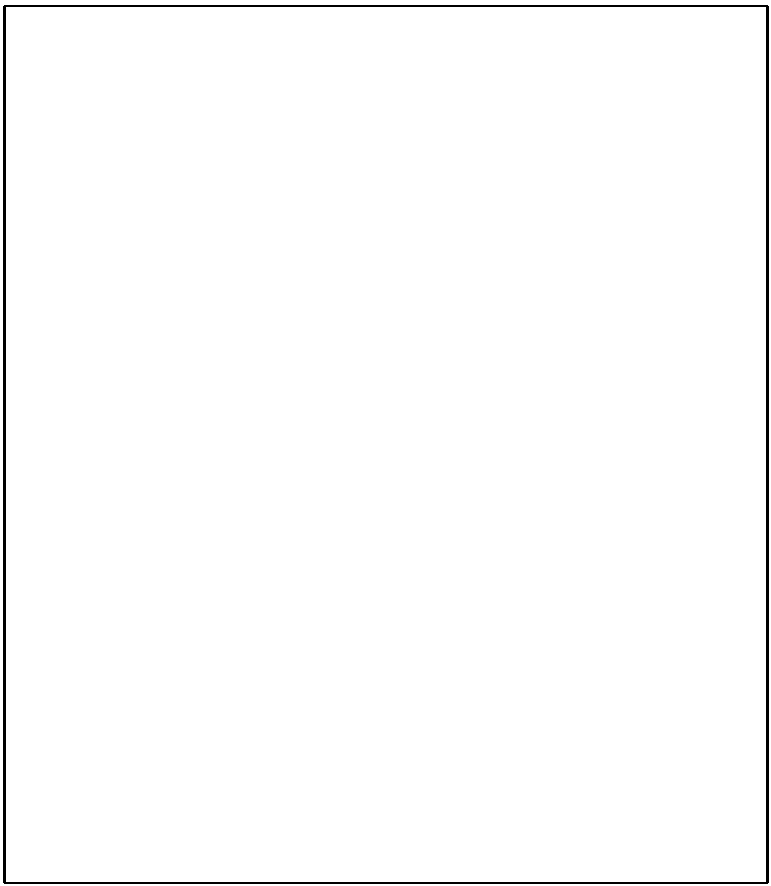

\centering
\begin{tabular}{|p{10.0cm}|}
\hline
\\ \\ \\ \\ \\ \\ \\ \\ \\ \\ \\ \\ \\ \\ \\ \\ \\
\\ \\ \\ \\ \\ \\ \\ \\ \\ \\ \\ \\ \\ \\ \\ \\ \\
\hline
\end{tabular}
\caption{The integral luminosity function in the 0.2-10 keV band of
  the X-ray sources in NGC~4261.  The solid line gives a broken
  power-law with fixed slopes, from Gilfanov (2004), renormalized to
  our data.  We overlay the single power-law (dashed line) with
  a slope $\alpha_3=-0.61$ as suggested by Grimm et al. (2003) to
  describe a population of HMXBs.}
\label{fig7}
\end{figure}

We verified the above statement by comparing the cumulative X-ray
luminosity function (XLF) of our sources with those derived from
observations of other galaxies (Figure \ref{fig7}).  In particular, we
use the functional forms derived by \cite{Grimm03} and
\cite{Gilfanov04} considered by these authors as a signature of a
universal behavior for the HMXB and the LMXB populations,
respectively; namely a single power-law for HMXB, with slope of
$\alpha =-0.61$, and a broken power-law for LMXB, with slopes of
$\alpha_1=-0.8,$ and $\alpha_2=-4$ for the high luminosity
end\footnote{These values are taken from Table 3 of
\cite{Gilfanov04}.}.  In testing the consistency with these universal
functions, we have derived the XLF using all sources in Table
\ref{tab:sources}, except for the AGN (source no. 29) and the 3
interlopers (sources no. 27, 43 and 53).  We performed a log-log
maximum likelihood fitting procedure for both HMXB and LMXB functional
forms assuming fixed slopes. We considered as free parameter the
normalization factors and, in the case of a broken power law, the
luminosity of the break.  The resulting XLFs are shown in Figure
\ref{fig7}.  It is unquestionable that a single power-law cannot
reproduce the LF of the sources in NGC~4261.  It is also apparent that
the LMXB broken-power law is a good representation of the data given
also the value of the break luminosity, that we have left as a free
parameter in the fitting procedure, and that is consistent with the
Gilfanov value of $L_b = 5.1 \cdot 10^{38} \ergs$.  This result is
consistent with the hypothesis that the X-ray sources in this galaxy
are associated to a population of LMXBs.  Note that we do not take
into account incompleteness at low luminosities \citep{Kim04};
however, we point out that at least at the high luminosity end (where
incompleteness should not significantly affect the XLF slope) the
luminosity function is much steeper than the XLF of the HMXBs.

\section{Azimuthal anisotropy in the globular cluster distribution}

ZHFM03 have shown with a Kolmogorov-Smirnov test that the spatial
distribution of the X-ray sources in NGC~4261 is anisotropic.  Using
our list of 50 sources (sources n.  29, 27, 43 and 53 excluded) we
confirm the spatial anisotropy over a scale of about $4'$ with very
high confidence (P$>99.9$\% from the Kolmogorov-Smirnov test).  In the
preceding sections we have shown that the properties of the X-ray
sources in NGC~4261 are consistent with a population of LMXBs.  This
however does not explain yet their spatial anisotropy, which is in
stark contrast with the smooth isophotal distribution of the optical
light.  Given the close association with the GC population, which we
find also has a peculiar distribution, we have considered the
possibility that the anisotropy is nothing but a manifestation of an
anisotropy seeded in the GC population.  In order to test this
hypothesis we have first done a statistical analysis on the spatial
distribution of the GC candidates.

\begin{figure}
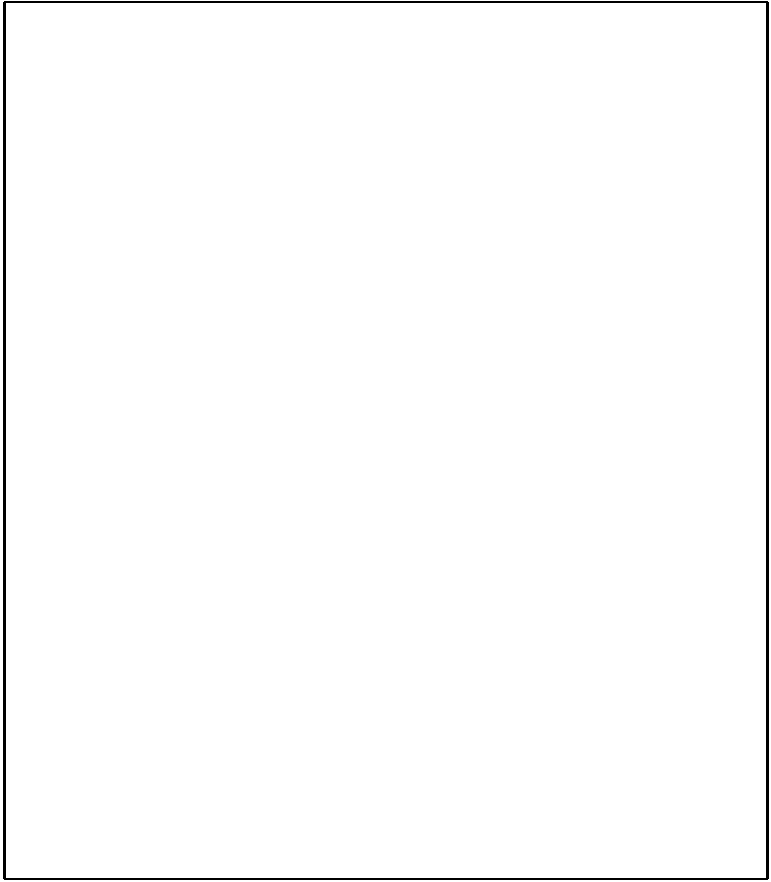

\centering 
\begin{tabular}{|p{10.0cm}|}
\hline
\\ \\ \\ \\ \\ \\ \\ \\ \\ \\ \\ \\ \\ \\ \\ \\ \\
\\ \\ \\ \\ \\ \\ \\ \\ \\ \\ \\ \\ \\ \\ \\ \\ \\
\hline
\end{tabular}
\caption{Iso-density contours from our adaptive kernel density
  analysis on the GC candidates from INT (thin line) and on the X-ray
  sources (thick line). The first contour represents the 2 $\sigma$
  level.  Note that the main over-densities in the distribution of
  X-ray sources coincide with those of GC candidates. }
\label{fig8}
\end{figure}

\begin{figure}
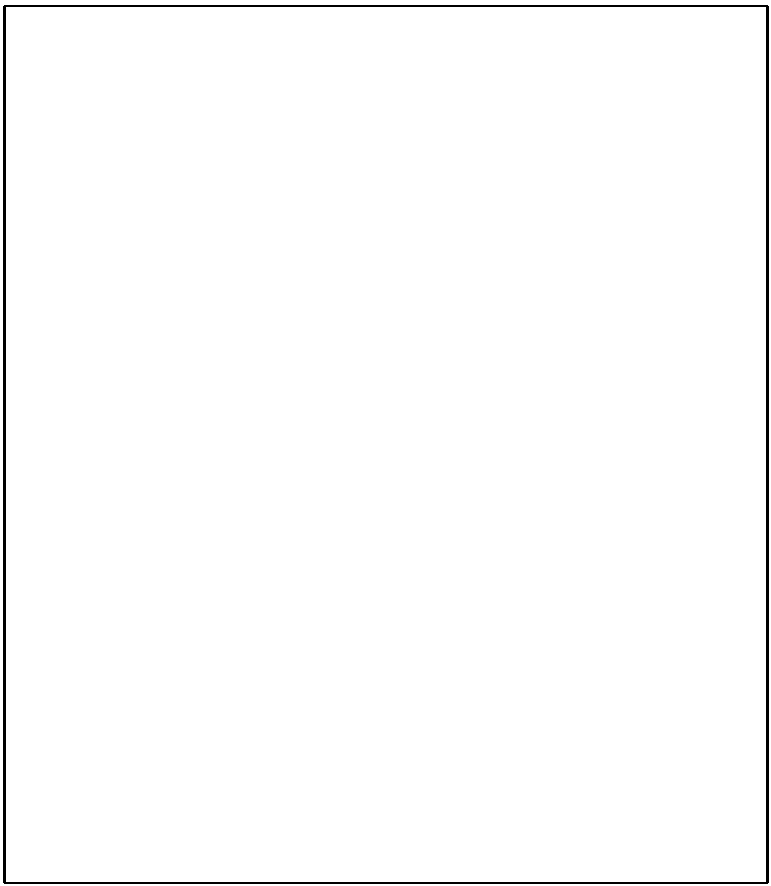

\centering 
\begin{tabular}{|p{10.0cm}|}
\hline
\\ \\ \\ \\ \\ \\ \\ \\ \\ \\ \\ \\ \\ \\ \\ \\ \\
\\ \\ \\ \\ \\ \\ \\ \\ \\ \\ \\ \\ \\ \\ \\ \\ \\
\hline
\end{tabular}
\caption{Left panel: Histogram of the surface number density (number
  of objects per square arcmin) of GC candidates (solid line) and of
  X-ray sources (dashed) as a function of their position in the field.
  Error bars are $\pm1\sigma$ computed according to \cite{gehrels96}.
  Right panel: The regions corresponding to each bin are indicated by
  squares superposed to the INT-model subtracted image.}
\label{fig9}
\end{figure}

In order to investigate both the details of the spatial distributions
of the X-ray and optical sources separately and their relation we have
compared the normalized surface density of sources again with an
adaptive kernel density analysis on the two sets of sources.  In
Figure \ref{fig8} we show the iso-intensity contours extracted from
the kernel analysis on the GC candidate sample (thin line) and on the
X-ray sources (thick line). With this figure as a guide, we have
defined 6 regions (see Figure \ref{fig9}, right panel) in which we
have simply computed the surface number density of sources for both
GCs and X-ray sources, which plot in Figure \ref{fig9} (left panel) as
a function of the region number.  The errors associated with each
entry reflects the total number of sources in the region, and gives a
direct measure of the significance of the observed region-to-region
variations, We also test the correlation between the two datasets
using the IRAF task \texttt{CROSSCOR} and we find that the highest
correlation between the two datasets is consistent with a null shift.

\section{Discussion and conclusions}

Our combined analysis of the X-ray and optical material available for
NGC~4261 shows that the large majority of the X-ray sources in
NGC~4261 are accreting LMXBs.  About 50\% of these sources have an
optical counterpart, that we identify with a GC.  Where color
information is available, the sources are associated with the red
(metal-rich) population, as already observed in several other
galaxies.  Further evidence of a LMXB origin comes from the
interpretation of the luminosity function, consistent with the
``universal'' distribution proposed by \cite{Gilfanov04}.

The photometric analysis of the GC population of NGC~4261 carried on a
portion of the galaxy ({\it HST}-SE field) shows that the properties
of the GCs are similar to those observed in other early-type galaxies
(\citealt{Cote98}; \citealt{Gebhardt99};
\citealt{West04}). Furthermore this similarity extends to the
connection between X-ray sources and GC population
(\citealt{Sarazin03}).  In particular we note that the X-ray sources
are preferentially associated with the brightest and reddest GCs. As
in other ellipticals, we find that the GCs hosting a X-ray source are
$\sim$7\% and that the X-ray sources associated with GCs are
$\sim$50\%, consistent with the result of \citealt{Kundu02}.  It
is therefore not surprising that NGC~4261 ranks among those galaxies
having a sizable fraction of X-ray sources in GCs.  The peculiarity of
NGC~4261 resides mainly in the large scale anisotropy of the X-ray
sources. Our adaptive kernel density analysis has highlighted the
presence of a peculiar spatial distribution in the GC candidate
population, and the existence of a correlation between the spatial
anisotropy of the X-ray sources and the main over-densities seeded in
the GC system.  To our knowledge, this is the first time that X-ray
sources indicate the presence of substructures in the distribution of
GC.

A non uniform spatial distribution of the GC system of NGC~4261 may
reflect a peculiar history of formation of this galaxy.  In the
paradigm of hierarchical clustering cosmogony, elliptical galaxies may
be the end result of binary or multiple major mergers among gas-rich
progenitors.  Major mergers likely redistribute the original GC
systems of the progenitor galaxies and trigger the formation of new
GCs in the center as well as along the tidal tails of the new object
arising from the merger. This would result in the coexistence of two
populations, metal rich and metal poor clusters, that have different
photometric properties, spatial distribution and dynamical
characteristics, as shown in a recent work by \cite{Li04}.
Hierarchical build-up also implies a second way by which galaxy can
grow, namely accretion of dwarf satellites that would lose their GCs
to the primary system. GCs coming from shredded satellites could also
contribute to spatial anisotropies and would be mostly metal-poor if
such satellites are similar to nearby dwarfs. Evidence that either of
these two mechanisms can account for a large fraction of the GC
systems in ellipticals is still lacking (\citealt{Strader04}).  A
close inspection of the photometric and kinematical properties of the
GCs can in principle help disentangle many aspects of such complex
galaxy genesis.  NGC~4261 could be an excellent target for further
investigations: combined Chandra and HST observations can thus
potentially shed light on how early-type galaxies were assembled. In
parallel, we are undertaking a theoretical study running
high-resolution simulations of galaxy mergers similar to those carried
on in \citealt{Kazantzidis04}, including a GC formation algorithm.

\acknowledgements 

We thank the anonymous Referee for useful comments that helped us to
improve the paper.  We thank Marco Scodeggio for providing us the
implementation of the algorithm used for the adaptive kernel density
analysis, and Michele Bellazzini for sharp comments and
suggestions. This work was supported by the Italian Ministry of
University and Research (MIUR) under the national program Prin 2003.

\clearpage

\clearpage

\clearpage

\clearpage

\clearpage

\clearpage

\clearpage

\clearpage

\clearpage

\clearpage

\begin{table}[!h]
\caption {Journal of all observations used in this work. SMAG indicates
the correspondent Johnson-Cousin filter.}
\[
\scriptsize
\begin{array}{p{0.08\linewidth}crrrrrrr}
\hline
\noalign{\smallskip}
\rm{Observation} & \rm{P.I.} & \rm{date} & \rm{instrument} & \rm{filters} & \rm{SMAG} & \rm{exposure \; time \; (s)} \\
\noalign{\smallskip}
\hline
\noalign{\smallskip}
optical & \rm{Ford} & \rm{13~Dec~1995} & \rm{{\it HST}} & \rm{F675W} & R & 2000 &  \\
optical & \rm{Ford} & \rm{13~Dec~1995} & \rm{{\it HST}} & \rm{F547M} & V & 800 &  \\
optical & \rm{Ford} & \rm{13~Dec~1995} & \rm{{\it HST}} & \rm{F791W} & I & 800 &  \\
optical & \rm{Sparks} & \rm{04~May~1995} & \rm{{\it HST}} & \rm{F702W} & R & 280 &  \\
optical & \rm{Osmond} & \rm{04-10~Feb~2000} & \rm{INT} & r' & R & 1500 &  \\
X-ray   & \rm{Birkinshaw} & \rm{06~May~2000} & \rm{ \it Chandra} & \rm{none} & - & 37210 &  \\
\noalign{\smallskip}
\hline
\end{array}
\]
\label{tab:obs}
\end{table}

\clearpage

\begin{table}
\caption {List of sources detected in the Field of NG~C4261. Count
  rates are in the full 0.3-7.0 keV band as given by $wavdetect$.
  X-ray luminosities are calculated assuming a power-law spectrum with
  an index of 1.7 and a distance of 32 Mpc. CPT indicates the
  association with a globular cluster (GC) in the HST field, a
  candidate globular cluster (GCC) from the INT list, or the name of
  the foreground/background galaxy. The question mark indicates that
  no optical coverage is available.}
\[
\scriptsize
\begin{array}{p{0.08\linewidth}crrrrrrr}
\hline
\noalign{\smallskip}
source & \rm{R.A.}  & \rm{Dec} & \rm{Count} \rm{Rate} & L_X(0.2-10 keV) & \rm{CPT} \\ 
number & \rm{(h \; m \; s)} & \rm{(h \; m \; s)} & (\times 10^{-4} \rm{s^{-1}}) & (\times 10^{38} \rm{s^{-1}}) & \\ 
\noalign{\smallskip}
\hline
\noalign{\smallskip}
1  & 12:19:14.44 & +05:48:48.12 & 3.76     \pm 1.22  & 4.93    &   \rm{ ... }   \\
2* & 12:19:16.81 & +05:49:50.43 & 10.65    \pm 1.97  & 13.96   &   \rm{ GCC }   \\
3  & 12:19:18.04 & +05:49:54.14 & 2.68     \pm 0.10  & 3.52    &   \rm{ GCC }   \\
4  & 12:19:18.23 & +05:49:12.10 & 2.03     \pm 0.89  & 2.67    &   \rm{ ... }   \\
5  & 12:19:18.67 & +05:47:43.23 & 4.15     \pm 1.22  & 5.45    &   \rm{ ... }   \\
6  & 12:19:19.18 & +05:48:49.31 & 2.63     \pm 0.10  & 3.45    &   \rm{ GCC }   \\
7  & 12:19:19.33 & +05:49:06.53 & 2.02     \pm 0.87  & 2.65    &   \rm{ ... }   \\
8  & 12:19:19.78 & +05:48:22.33 & 4.03     \pm 1.22  & 5.28    &   \rm{ ... }   \\
9  & 12:19:19.91 & +05:48:50.90 & 4.21     \pm 1.22  & 5.52    &   \rm{ ... }   \\
10 & 12:19:20.27 & +05:49:08.58 & 7.88     \pm 1.69  & 10.33   &   \rm{ GCC }   \\
11 & 12:19:20.44 & +05:48:02.72 & 5.18     \pm 1.37  & 6.79    &   \rm{ ... }   \\
12 & 12:19:20.60 & +05:48:35.89 & 4.97     \pm 1.37  & 6.52    &   \rm{ GC  }   \\
13 & 12:19:20.66 & +05:49:29.10 & 2.73     \pm 0.10  & 3.58    &   \rm{ GCC }   \\
14 & 12:19:20.93 & +05:49:27.09 & 5.83     \pm 1.46  & 7.64    &   \rm{ GCC }   \\
15 & 12:19:21.00 & +05:48:44.07 & 3.98     \pm 1.22  & 5.22    &   \rm{ GC  }   \\
16 & 12:19:21.07 & +05:50:17.38 & 3.26     \pm 1.12   & 4.27    &   \rm{ ... }   \\
17 & 12:19:21.08 & +05:48:07.21 & 2.36     \pm 0.94   & 3.09    &   \rm{ ... }   \\
18 & 12:19:21.20 & +05:47:52.14 & 5.55     \pm 1.46   & 7.27    &   \rm{ GCC }   \\
19 & 12:19:22.00 & +05:49:47.02 & 3.43     \pm 1.12   & 4.50    &   \rm{ ?   }   \\
20 & 12:19:22.16 & +05:50:59.57 & 3.07     \pm 1.06   & 4.02    &   \rm{ ... }   \\
21 & 12:19:22.35 & +05:48:47.95 & 4.27     \pm 1.27   & 5.60    &   \rm{ GC  }   \\
22 & 12:19:22.54 & +05:50:17.00 & 5.75     \pm 1.46   & 7.54    &   \rm{ GCC }   \\
23 & 12:19:22.67 & +05:47:23.04 & 3.69     \pm 1.17   & 4.84    &   \rm{ ... }   \\
24 & 12:19:22.81 & +05:46:44.62 & 3.45     \pm 1.12   & 4.52    &   \rm{ ... }   \\
25 & 12:19:22.99 & +05:51:11.50 & 1.60     \pm 0.79   & 2.10    &   \rm{ ... }   \\
26 & 12:19:23.00 & +05:50:58.55 & 5.53     \pm 1.41   & 7.25    &   \rm{ ... }   \\
27 & 12:19:23.10 & +05:47:41.43 & 7.16     \pm 1.66   & 9.38    &   \rm{VCC~344}   \\
28 & 12:19:23.20 & +05:48:07.16 & 2.69     \pm 0.10   & 3.53    &   \rm{ ... }   \\
29 & 12:19:23.21 & +05:49:29.76 & 1249.41  \pm 21.02  & 1638.28 &   \rm{ AGN }   \\
30 & 12:19:23.23 & +05:50:13.44 & 3.44     \pm 1.12   & 4.51    &   \rm{ GCC }   \\
31 & 12:19:23.50 & +05:49:36.06 & 11.02    \pm 4.07   & 14.45   &   \rm{ ?   }   \\
32 & 12:19:23.61 & +05:47:55.54 & 4.70     \pm 1.32   & 6.16    &   \rm{ GC  }   \\
33 & 12:19:24.01 & +05:49:26.42 & 9.92     \pm 3.02   & 13.00   &   \rm{ ?   }   \\
34 & 12:19:24.25 & +05:47:54.75 & 5.92     \pm 1.50   & 7.76    &   \rm{ ... }   \\
35 & 12:19:24.60 & +05:51:04.86 & 8.18     \pm 1.73   & 10.73   &   \rm{ GCC }   \\
36 & 12:19:24.80 & +05:50:57.37 & 2.40     \pm 0.94   & 3.15    &   \rm{ ... }   \\
37*& 12:19:24.83 & +05:50:08.61 & 3.69     \pm 1.17   & 4.84    &   \rm{ GC  }   \\
38 & 12:19:24.93 & +05:49:32.11 & 5.50     \pm 1.41   & 7.22    &   \rm{ GC  }   \\
39 & 12:19:25.50 & +05:47:41.85 & 6.53     \pm 1.54   & 8.57    &   \rm{ GCC }   \\
40 & 12:19:25.55 & +05:49:46.17 & 4.92     \pm 1.41   & 6.45    &   \rm{ GC  }   \\
41*& 12:19:25.65 & +05:50:28.75 & 2.73     \pm 1.06   & 3.58    &   \rm{ GC  }   \\
42 & 12:19:25.95 & +05:52:20.50 & 3.96     \pm 1.22   & 5.19    &   \rm{ ... }   \\
43*& 12:19:26.06 & +05:50:13.69 & 2.76     \pm 0.10   & 3.62    &   \rm{ galaxy }   \\
44 & 12:19:26.89 & +05:50:44.15 & 5.74     \pm 1.46   & 7.52    &   \rm{ GC  }   \\
45 & 12:19:29.22 & +05:50:00.59 & 2.91     \pm 1.06   & 3.82    &   \rm{ GCC }   \\
46 & 12:19:29.78 & +05:51:14.35 & 7.46     \pm 1.66   & 9.78    &   \rm{ ... }   \\
47 & 12:19:29.97 & +05:50:28.20 & 5.64     \pm 1.46   & 7.39    &   \rm{ ... }   \\
48 & 12:19:30.51 & +05:50:07.97 & 3.11     \pm 1.06   & 4.08    &   \rm{ ... }   \\
49 & 12:19:31.08 & +05:50:07.96 & 2.50     \pm 0.10   & 3.28    &   \rm{ GCC }   \\
50 & 12:19:31.98 & +05:51:54.46 & 5.60     \pm 1.46   & 7.35    &   \rm{ GCC }   \\
51 & 12:19:32.08 & +05:51:49.47 & 4.85     \pm 1.41   & 6.35    &   \rm{ ... }   \\
52 & 12:19:33.93 & +05:50:35.75 & 5.65     \pm 1.46   & 7.41    &   \rm{ ... }   \\
53 & 12:19:35.52 & +05:50:48.93 & 4.04     \pm 1.27   & 5.30    &   \rm{NGC~4264}   \\
54 & 12:19:36.87 & +05:51:56.22 & 4.44     \pm 1.32   & 5.82    &   \rm{ ... }   \\
\noalign{\smallskip}
\hline
\end{array}
\]
\label{tab:sources}
\end{table}

\clearpage

\end{document}